\documentclass[pre,preprint,showpacs,amsmath,amssymb]{revtex4}

\usepackage{wasysym}
\usepackage{graphicx}
\usepackage{dcolumn}
\usepackage{bm}

\begin{document}

\title{Effect of interactions on the diffusive expansion of a Bose-Einstein condensate in a 3D random potential\\}

\author{N. Cherroret and S. E. Skipetrov}
\affiliation{
Universit\'{e} Joseph Fourier, Laboratoire de Physique et Mod\'{e}lisation des Milieux Condens\'{e}s, CNRS, 25 rue des Martyrs, BP 166, 38042 Grenoble, France
}

\date{\today}

\begin{abstract}
We theoretically study the influence of weak interactions on the diffusive expansion of a Bose-Einstein condensate in a three-dimensional random potential. For this purpose we develop a perturbative approach and calculate analytically the first-order nonlinear correction to the ensemble-averaged atomic density as a function of position and time. We find that interactions renormalize the typical diffusion coefficient of the condensate. The magnitude of the nonlinear correction is controlled  by a single dimensionless parameter that includes both interaction and disorder strengths.
\end{abstract}


\maketitle
\section{Introduction}

Ultracold atomic gases have proven to be a powerful tool to revisit many phenomena of mesoscopic physics. In particular, the study of Bose-Einstein condensates (BECs) in random potentials allows better understanding of the interplay between disorder and interactions \cite{reviewBEC}. The main advantage of experiments with BECs is their great versatility. In particular, the strength of the interatomic interaction can be tuned almost at will, making use of Feshbach resonances. In the same way, the realization of optical speckle patterns allows an unprecedented control of the parameters of disorder. A prominent experimental  example of this is the observation of first hints suggesting a Bose-glass phase for strongly interacting BECs in disordered potentials \cite{Fallani07}. In the last few years, extensive experimental and theoretical works have also been carried out on the expansion of BECs in one-dimensional (1D) random potentials \cite{experimental_papers,theoretical_papers}, with particular interest in the phenomenon of Anderson localization \cite{Billy, Roati,Sanchez-Palencia}. However up to now, very few works have concerned the expansion of BECs in three-dimensional (3D) potentials. In three dimensions the situation is very rich since unlike in 1D, a critical energy separates localized from extended states \cite{Anderson}. Recently, a systematic study of Anderson localization of BECs based on the self-consistent theory of Vollhardt and W\"{o}lfle has been proposed \cite{Sergey}. Somewhat earlier, the diffusive expansion of a BEC in a 3D random potential, which predominates when the disorder is weak enough, has been studied by Shapiro \cite{Shapiro}. In these works however, the role of interactions that come into play during the expansion of the BEC in the random potential has been ignored. At long times, when the condensate is sufficiently dilute, this assumption is known to be true in the absence of disorder \cite{Pitaevskii}, and seems natural for weak disorder.  In the latter case it has been validated by numerical simulations in 1D \cite{Sanchez-Palencia}. However, even in 1D, no consensus on the effect of interactions on the expansion of BECs exists \cite{Sanchez-Palencia,Pikivsky08}, and, to our knowledge, there is neither analytical nor numerical work about this issue in 3D.

In this paper we propose the first analytical study of the nonlinear corrections due to interatomic interactions to the density profile of a BEC expanding in a 3D random potential. We assume that the disorder is weak and neglect Anderson localization. Therefore, we are interested in the effect of interactions on the \emph{diffusive} expansion of the condensate. In the absence of interactions the density averaged over an ensemble of realizations of the random potential is expected to decay as $(D_\mu t)^{-3/2}$ at long times, where $D_\mu$ is the diffusion coefficient of a particle of energy $\mu$, the chemical potential of the condensate \cite{Shapiro}. Surprisingly, we find that the first-order nonlinear correction follows the same characteristic decay. Interactions thus simply renormalize the diffusion coefficient of the condensate $D_\mu$, at least at the first-order of perturbation theory. As expected, we find that repulsive interactions reinforce the diffusion unlike Anderson localization which inhibits it. Interestingly, our results show that nonlinear effects get stronger as disorder increases.

\section{Perturbative treatment}\label{Perturbative}

Consider a weakly interacting Bose-Einstein condensate of $N\gg 1$ atoms of mass $m$ expanding in a three-dimensional random potential $V(\textbf{r})$, for which we assume a white-noise gaussian statistics: $\overline{V(\textbf{r})V(\textbf{r}^\prime)}=\hbar^4\pi/m^2\ell\delta(\textbf{r}-\textbf{r}^\prime)$, where $\ell$ is the mean free path \cite{Akkermans} and the overline denotes averaging over realizations of the random potential. We treat the condensate in the framework of the mean field Gross-Pitaevskii equation
\begin{equation} \label{Pitaevskii}
i \hbar \dfrac{\partial\psi}{\partial t}=\left[-\dfrac{\hbar^2}{2m}\nabla^2+V(\textbf{r})+g|\psi|^2\right]\psi,
\end{equation}
where $\psi(\textbf{r},t)$ is the condensate wave function and $g=4\pi\hbar^2a/m$ is the strength of interactions, with $a$ the scattering length \cite{Pitaevskii}. Note that the following reasoning is valid for both repulsive ($a>0$) and attractive ($a<0$) interactions. By Fourier transforming Eq. (\ref{Pitaevskii}) with respect to time we obtain

\begin{equation} \label{Pitaevskii_Fourier}
\left[-\epsilon-\dfrac{\hbar^2}{2m}\nabla^2+V(\textbf{r})\right]\psi_{\epsilon}(\textbf{r})+g\int_{-\infty}^{\infty}\dfrac{d\epsilon_1}{2\pi}\int_{-\infty}^{\infty}\dfrac{d\epsilon_2}{2\pi}\psi_{\epsilon_1}(\textbf{r})\psi_{\epsilon_2}^*(\textbf{r})\psi_{\epsilon-\epsilon_1+\epsilon_2}(\textbf{r})=0,
\end{equation}
where $\psi_{\epsilon}(\textbf{r})=\int dt \psi(\textbf{r},t)\exp(i\epsilon t/\hbar)$. To describe the effect of weak interactions ($|g|\ll1$) on the expansion of the condensate we make use of perturbation theory and write $\psi_\epsilon(\textbf{r})=\psi^{(0)}_\epsilon(\textbf{r})+\delta\psi_\epsilon(\textbf{r})$, where $|\delta\psi_\epsilon|\ll|\psi_\epsilon^{(0)}|$. A precise condition of validity of this approach will be given later on (see section \ref{correction}). In the following, the superscript $^{(0)}$ will refer to quantities free of interactions. Inserting this expression into Eq. (\ref{Pitaevskii_Fourier}) and keeping only the lowest-order terms in $\delta\psi_\epsilon$ and $g$ we find that $\psi^{(0)}_\epsilon$ obeys the linear Shr\"{o}dinger equation

\begin{equation} \label{Shrodinger}
\left[-\epsilon-\dfrac{\hbar^2}{2m}\nabla^2+V(\textbf{r})\right]\psi^{(0)}_{\epsilon}(\textbf{r})=0,
\end{equation}
whereas the first-order term $\delta\psi_\epsilon$ satisfies
\begin{equation} \label{Shrodinger_g}
\left[-\epsilon-\dfrac{\hbar^2}{2m}\nabla^2+V(\textbf{r})\right]\delta\psi_{\epsilon}(\textbf{r})=g\int_{-\infty}^{\infty}\dfrac{d\epsilon_1}{2\pi}\int_{-\infty}^{\infty}\dfrac{d\epsilon_2}{2\pi}\psi^{(0)}_{\epsilon_1}(\textbf{r})\psi^{(0)*}_{\epsilon_2}(\textbf{r})\psi^{(0)}_{\epsilon-\epsilon_1+\epsilon_2}(\textbf{r}).
\end{equation}
Eqs. (\ref{Shrodinger}) and (\ref{Shrodinger_g}) form the basis for studying the behavior of the time-dependent ensemble-averaged atomic density $\bar{n}(\textbf{r},t)=\overline{|\psi(\textbf{r},t)|^2}$. By expanding $\bar{n}$ and keeping only the lowest terms in $\delta\psi$ we readily obtain
\begin{equation}\label{n}
\bar{n}(\textbf{r},t)\simeq\bar{n}^{(0)}(\textbf{r},t)+\overline{\Delta n}(\textbf{r},t),
\end{equation}
where $\bar{n}^{(0)}(\textbf{r},t)=\overline{|\psi^{(0)}(\textbf{r},t)|^2}$ is the atomic density in the absence of interactions and $\overline{\Delta n}(\textbf{r},t)=2{\rm Re}(\overline{\psi^{(0)*}(\textbf{r},t)\delta\psi(\textbf{r},t)})$ is the first-order correction to $\bar{n}$, with $\psi^{(0)}(\textbf{r},t)$ and $\delta\psi(\textbf{r},t)$ the Fourier transforms of $\psi^{(0)}_\epsilon(\textbf{r})$ and $\delta\psi_\epsilon(\textbf{r})$ respectively. The two next sections are devoted to the separate study of $\bar{n}^{(0)}$ and $\overline{\Delta n}$.

\section{Expansion in the absence of interactions}\label{Diffusive}

\subsection{Diffusive expansion}
In this section we remind a few results concerning the averaged density profile $\bar{n}^{(0)}(\textbf{r},t)$ free of interactions in three dimensions. We assume that the condition of weak disorder $k_\mu\ell\gg 1$ \cite{Akkermans} is fulfilled, where $k_\mu$ is the wavevector at energy $\mu$. As a consequence, the condensate expands mostly by diffusion in the random potential \cite{Sergey,Shapiro}. This point will be discussed more deeply in section \ref{Link}. In the absence of interactions, the wave function of the condensate at time $t$ and position $\textbf{r}$ is given by
\begin{equation}\label{psi0}
\psi^{(0)}(\textbf{r},t)=\int\dfrac{d\epsilon}{2\pi}\int d^3\textbf{r}^\prime G_\epsilon(\textbf{r},\textbf{r}^\prime)\phi(\textbf{r}^\prime)e^{-i\epsilon t/\hbar},
\end{equation}
where $\phi$ is the initial condensate wave function and $G_\epsilon$ is the Green's function of Eq. (\ref{Shrodinger}). From here on, we consider large distances $r\gg \ell$ and long times $t\gg(\ell/v_\mu)$, where $v_\mu=\hbar k_\mu/m$ is the velocity of a particle with kinetic energy $\mu$. The mean free time $\ell/v_\mu$ can be regarded as the time separating the ballistic regime $t\lesssim\ell/v_\mu$  from the multiple scattering one $t\gg\ell/v_\mu$ where atoms are scattered many times on the random potential. Under these assumptions we derive from Eq. (\ref{psi0}) the following expression for $\overline{n}^{(0)}$ (more details can be found for example in Ref. \cite{Shapiro}):
\begin{equation} \label{atomic_density}
\bar{n}^{(0)}(\textbf{r},t)=\int\dfrac{d^3\textbf{k}}{(2\pi)^3}|\phi(\textbf{k})|^2P_{\epsilon_k}(\textbf{r},t).
\end{equation}
Here $\epsilon_k=\hbar^2k^2/2m$ and $P_{\epsilon}(\textbf{r},t)=\exp(-|\textbf{r}-\textbf{r}^\prime|^2/4D_\epsilon t)/(4\pi D_\epsilon t)^{3/2}$ is the diffusion propagator in three dimensions for a particle at energy $\epsilon$. $D_\epsilon=(1/3)\ell v_\epsilon$ is the diffusion coefficient of a particle of velocity $v_\epsilon=\sqrt{2\epsilon/m}$ \cite{Akkermans}. The momentum distribution of the condensate $|\phi(\textbf{k})|^2\propto (1-k^2/2k_\mu^2)H(1-k/\sqrt{2}k_\mu)$, with $H(x)$ the Heaviside step function, is obtained from the following expansion scenario \cite{Sanchez-Palencia,Sergey,Shapiro}: the BEC is generally produced in a harmonic trap potential of frequency $\omega$. The trap is turned off at some time, which causes the rapid expansion of the condensate, driven by strong interatomic interactions. At this initial stage of expansion we neglect the effect of disorder. After a time of the order of $1/\omega$, interactions become weak and the momentum distribution is given by $|\phi(\textbf{k})|^2$ \cite{Kagan}. We choose our initial time $t=0$ within this latter stage. 

\begin{figure}[h]
\includegraphics[width=11cm]{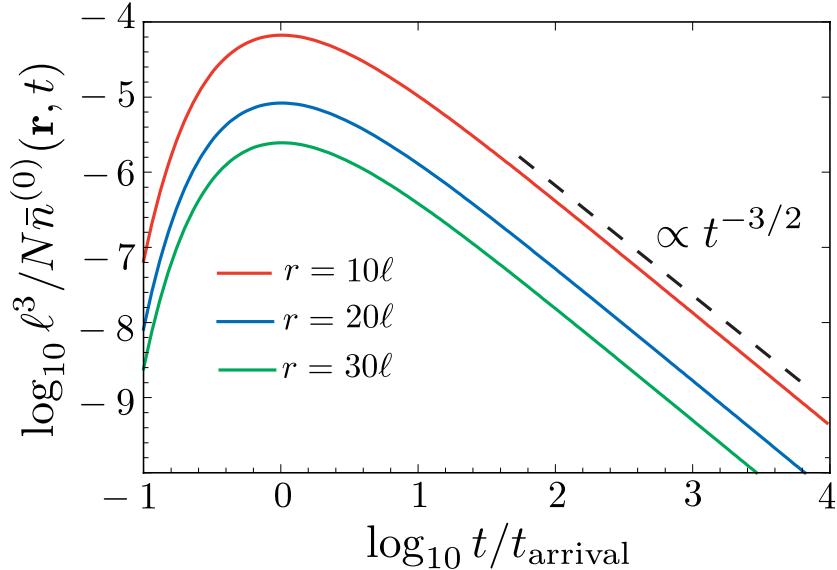}
\caption{\label{n0_diffusion} 
(color online). Ensemble-averaged atomic density $\bar{n}^{(0)}(\textbf{r},t)$ of a BEC expanding without interactions in a 3D random potential, in the regime of weak disorder $k_\mu\ell\gg 1$, as a function of time and for three different distances $r=|\textbf{r}|$ from the initial location of the condensate. The dashed line is a $1/t^{3/2}$ asymptote. The time scale is in units of $t_\text{arrival}$ such that all maxima fall in the same abscissa.
}  
\end{figure}

The calculation of the integral in Eq. (\ref{atomic_density}) is straightforward and leads to:
\begin{equation}\label{atomic_density_explicit}
\bar{n}^{(0)}(\textbf{r},t)=\dfrac{N}{(D_\mu t)^{3/2}}f\left(\dfrac{r^2}{\sqrt{D_\mu t}}\right),
\end{equation}
where $f(x)$ can be expressed through special functions and $f(x) \simeq 0.04$ for $x \ll 1$. As pointed out in \cite{Shapiro}, the density profile is driven by a single parameter $D_\mu=(1/3)v_\mu\ell$, which can be regarded as the typical diffusion coefficient of the condensate. We plot the density profile in Fig. \ref{n0_diffusion} as a function of time for three different distances $r$. $\bar{n}^{(0)}(\textbf{r},t)$ exhibits the well known decay in $t^{-3/2}$ in the long time limit $t\gg t_{\text{arrival}}$, where $t_{\text{arrival}}\sim r^2/D_\mu\sim(\ell/v_\mu)(r/\ell)^2$ is the ``arrival "time at which the density reaches the maximum.

\subsection{Anderson localization}\label{Link}

\begin{figure}[h]
\includegraphics[width=11cm]{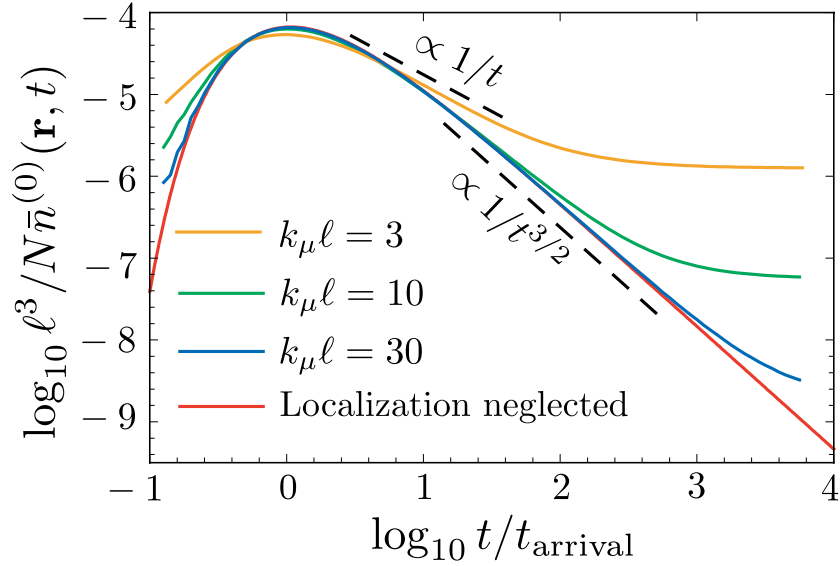}
\caption{\label{n0_diffloc} 
(color online). Ensemble-averaged atomic density $\bar{n}^{(0)}(\textbf{r},t)$ of a BEC expanding without interactions in a 3D random potential, for three different values of the disorder parameter $k_\mu\ell$ and for $r=10\ell$ (orange, green and blue upper curves). Here unlike Fig. \ref{n0_diffusion}, the effects of Anderson localization are fully taken into account. For moderate values of $k_\mu\ell$, the curves exhibit first a $1/t$ decay and then saturation at some constant value, in agreement with Ref.\cite{Sergey}. When $k_\mu\ell\gg 1$, the atomic density exhibits a $1/t^{3/2}$ decay, before saturating at very long times as well. The red lower curve is obtained from Eq. (\ref{atomic_density_explicit}), i.e. by neglecting Anderson localization effects.
}  
\end{figure}

It must be noted that Eq. (\ref{atomic_density_explicit}) describes a purely diffusive expansion and does not include the effects of Anderson localization. In Ref. \cite{Sergey}, the authors found $\bar{n}^{(0)}\propto 1/t$ for $t>t_\text{arrival}$. A question then naturally arises: to which extent the diffusive result (\ref{atomic_density_explicit}) is correct? In Ref. \cite{Sergey} the density profile has been studied for moderate values of $k_\mu\ell$, that is in the regime of strong disorder. When $k_\mu\ell$ is increased, the $1/t$ decay is expected to cross over to the $1/t^{3/2}$ decay typical for diffusion. To illustrate this cross-over we plot  $\bar{n}^{(0)}(\textbf{r},t)$ including localization effects in Fig. \ref{n0_diffloc} for three values of $k_\mu\ell$ and for $r=10\ell$ (three upper curves, orange, green and blue online). The lower curve in Fig. \ref{n0_diffloc}  (red online) is obtained from Eq. (\ref{atomic_density_explicit}), i.e. by neglecting Anderson localization effects.
For $k_\mu\ell=3$, a significant fraction of atoms of the condensate are localized. At very short times, all diffusive atoms have already flown away and $\bar{n}^{(0)}\propto 1/t$, in agreement with Ref. \cite{Sergey}. For weak disorder ($k_\mu\ell=30 in Fig. \ref{n0_diffloc}$), the situation is very different: most atoms of the condensate propagate diffusively. Therefore the atomic density exhibits a $1/t^{3/2}$ decay characteristic for diffusion. Whatever $k_\mu\ell$ however, at very long times the atomic density saturates at a constant value $\bar{n}^{(0)}_\text{loc}\sim h(k_\mu\ell)(N/r^3)(\ell/r)^{1/\nu}$, where $\nu$ is the critical exponent of the localization transition, and $h(k_\mu\ell)\propto 1/(k_\mu\ell)^3$ for $k_\mu\ell\gg 1$ and $h(k_\mu\ell)\propto \text{const}$ for $k_\mu\ell\ll 1$ \cite{Sergey}. We can estimate the time after which the diffusive approach breaks down by requiring that the saturation value $\bar{n}^{(0)}_{\text{loc}}$ is smaller than the limit $\bar{n}^{(0)}_{\text{dif}}\sim N/(D_\mu t)^{3/2}$ of Eq. (\ref{atomic_density_explicit}) at long times $t\gg t_{\text{arrival}}$. This yields
\begin{equation}\label{tloc}
t_{\text{loc}}\sim(\ell/v_\mu)(r/\ell)^{2(1+1/3\nu)}(k_\mu\ell)^2,
\end{equation}
as the maximum time at which diffusion model can be applied. When $k_\mu\ell\gg 1$ and $r\gg\ell$, $t_\text{loc}$ is always larger than $t_\text{arrival}$ and the diffusion model is hence valid in a broad time interval. To sum up, the condensate behaves diffusively ($\bar{n}^{(0)}\propto t^{-3/2}$) until $t=t_\text{loc}$, and beyond localization effects start to ``freeze" the density profile at $n^{(0)}=\bar{n}^{(0)}_\text{loc}$. The conditions $k_\mu\ell\gg 1$ and $t\ll t_\text{loc}$ will be assumed throughout the remainder of the paper.
\section{Nonlinear correction to the atomic density}\label{Correction}
In this section we are interested in the first-order nonlinear correction $\overline{\Delta n}$ to the density profile [see Eq. (\ref{n})]. We first discuss the effect of interactions on the mean free path, and then develop a diagrammatic approach for calculating $\overline{\Delta n}$.

\subsection{Mean free path}\label{mfp}
It should be noted that the mean free path $\ell$ is in principle modified in the presence of interactions. However in the following we will neglect the nonlinear corrections to $\ell$. This approximation is legitimate in the limit of long times, as we now show by making use of a criterion initially developed in \cite{Spivak} for monochromatic classical waves. 

The mean free path is weakly affected by interactions provided that the scattering on the effective potential $g|\psi(\textbf{r},t)|^2$ is weak as compared to the scattering on the random potential. In other words, the mean free path $\ell^{(1)}$ associated with the interaction potential should be much larger than the mean free path $\ell^{(0)}=\ell$ associated with the random potential. $\ell^{(1)}$ can be evaluated from the short-range $``C_1"$ correlation function of the density $\overline{\delta n(\textbf{r},t)\delta n(\textbf{r}^\prime,t)}$ which has been recently studied by Henseler and Shapiro \cite{Henseler}. We neglect here the contribution due to the long-range $``C_2"$ part of the density correlation function. This would be questionable for a condensate propagating in a quasi-one dimensional waveguide where $C_2$ was predicted to dominate at long times \cite{Nicolas08}. However the situation is different in an unbounded 3D random potential where we expect $C_2$ to decay with time. Indeed, let us consider a quasi-monochromatic wave pulse emitted from some point in an unbounded three-dimensional  disordered medium. A straightforward calculation similar to that of Ref. \cite{Nicolas08} yields $C_2\propto(1/t^2)/(k\ell)^2$ in the limit of long times. $C_2$ is therefore negligible with respect to $C_1$.

From \cite{Henseler} we write $\overline{\delta n(\textbf{r},t)\delta n(\textbf{r}^\prime,t)}\simeq (2\pi\ell/k_\mu^2)\bar{n}^{(0)}(\textbf{r},t)^2\delta{(\textbf{r}-\textbf{r}^\prime)}$. For times $t\gg t_{\text{arrival}}$, this leads to $\ell^{(1)}\simeq (\hbar^4\pi/m^2)(k_\mu^2/2\pi\ell)(D_\mu t)^3/(g^2N^2)$. The condition $\ell^{(1)}\gg\ell$ then gives
\begin{equation}\label{t1}
t\gg t_1=p^{2/3}\left(\dfrac{\ell}{v_\mu}\right)\dfrac{1}{(k_\mu\ell)^{1/3}},
\end{equation}
where we introduced a dimensionless parameter $p=aN/\ell\sqrt{k_\mu\ell}$ that includes both interaction and disorder strengths. At this point, it is sufficient to keep in mind that in the weak disorder limit and under conditions of typical experiments, $p\ll 1$. For example with the data from Refs. \cite{experimental_papers} and \cite{Billy} we find $p\sim10^{-2}$. Therefore, the characteristic time $t_1$ is much smaller than $\ell/v_\mu$, such that the criterion (\ref{t1}) is automatically satisfied for long times $t\gg\ell/v_\mu$.

\subsection{Diffusion coefficient}\label{correction}

\begin{figure}[h]
\includegraphics[width=9cm]{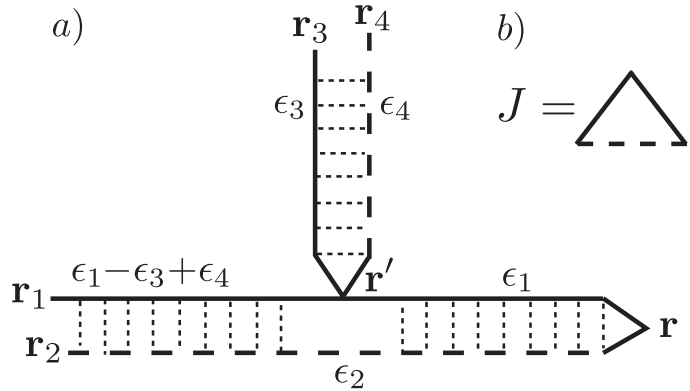}
\caption{\label{diagram} 
a) Diagram for the kernel $K$ giving the first nonlinear correction to the atomic density. The solid and dashed lines represent Green's functions $G_{\epsilon_i}$ and  $G^*_{\epsilon_j}$, respectively, where $\epsilon_i=\epsilon_1$, $\epsilon_3$ or $\epsilon_1-\epsilon_3+\epsilon_4$ and $\epsilon_j=\epsilon_2$ or $\epsilon_4$. The two parallel $G$ lines connected by dotted ``ladders" symbolize averages of products of two Green's functions, $\overline{G_{\epsilon_i}G^*_{\epsilon_j}}$. b) Diagram J of Eq. (\ref{deltapsi_implicit2}).
}  
\end{figure}

We now consider $\overline{\Delta n}(\textbf{r},t)$. From Eq. (\ref{Shrodinger_g}) we obtain
\begin{equation}\label{deltapsi}
\delta\psi(\textbf{r},t)=\int\dfrac{d\epsilon}{2\pi}\int d^3\textbf{r}^\prime G_\epsilon(\textbf{r},\textbf{r}^\prime)S(\textbf{r}^\prime)e^{-i\epsilon t/\hbar},
\end{equation}
where $S(\textbf{r}^\prime)$ is the right-hand side of Eq. (\ref{Shrodinger_g}) evaluated at $\textbf{r}^\prime$. Combining Eqs. (\ref{psi0}) and (\ref{deltapsi}) we obtain

\begin{eqnarray}\label{deltapsi_implicit}
\overline{\Delta n}(\textbf{r},t)&=&2{\rm Re}\left[\overline{\psi^{(0)*}(\textbf{r},t)\delta\psi(\textbf{r},t)}\right]\nonumber \\
&=&2{\rm Re}\bigg[\dfrac{g}{(2\pi)^4}\int\prod_{j=1}^4d\epsilon_jd^3\textbf{r}_jd^3\textbf{r}^\prime e^{(-i/\hbar)(\epsilon_1-\epsilon_2)t)}K(\textbf{r},t;\textbf{r}^\prime,\{\textbf{r}_j\},\{\epsilon_j\})\nonumber\\
&&\times\phi(\textbf{r}_1)\phi^*(\textbf{r}_2)\phi(\textbf{r}_3)\phi^*(\textbf{r}_4)\bigg],
\end{eqnarray}
where the six-point kernel $K$ is given by the connected part of a product of five Green's functions, averaged over disorder:
\begin{equation}\label{Kernel}
K=\overline{G_{\epsilon_1}^{}(\textbf{r},\textbf{r}^\prime)G_{\epsilon_2}^*(\textbf{r},\textbf{r}_2)G_{\epsilon_3}^{}(\textbf{r}^\prime,\textbf{r}_3)G_{\epsilon_4}^*(\textbf{r}^\prime,\textbf{r}_4)G_{\epsilon_1-\epsilon_3+\epsilon_4}^{}(\textbf{r}^\prime,\textbf{r}_1)}.
\end{equation}
The problem therefore reduces to finding the largest contribution to the kernel $K$ in the weak disorder limit $k_\mu\ell\gg 1$. This contribution is given by the diagram depicted in Fig. \ref{diagram}a. The diagram describes interaction of two matter waves at a point $\textbf{r}^\prime$. Diagrams of this type have been introduced some time ago for monochromatic classical waves in the context of  nonlinear optics of disordered media \cite{Agranovich91}. In our case the situation is more general since $K$ contains a product of Green's functions at five different energies.
Writing down explicitly the diagram of Fig. \ref{diagram}a, we obtain
\begin{eqnarray}\label{deltapsi_implicit2}
\overline{\Delta n}(\textbf{r},t)&=&
2{\rm Re}\bigg[\dfrac{gJ}{(2\pi)^4}\left(\dfrac{\hbar v_\mu}{2\pi\rho\ell}\right)^2\int\prod_{j=1}^4d\epsilon_jd^3\textbf{r}_jd^3\textbf{r}^\prime e^{(-i/\hbar)(\epsilon_1-\epsilon_2)t}\phi(\textbf{r}_1)\phi^*(\textbf{r}_2)\phi(\textbf{r}_3)\phi^*(\textbf{r}_4) \nonumber\\
&& \times\overline{G_{\epsilon_1}(\textbf{r},\textbf{r}^\prime)G^*_{\epsilon_2}(\textbf{r},\textbf{r}^\prime)}\times \overline{G_{\epsilon_3}(\textbf{r}^\prime,\textbf{r}_3)G^*_{\epsilon_4}(\textbf{r}^\prime,\textbf{r}_4)}\times \overline{G_{\epsilon_1-\epsilon_3+\epsilon_4}(\textbf{r}^\prime,\textbf{r}_1)G^*_{\epsilon_2}(\textbf{r}^\prime,\textbf{r}_2)}
\bigg],
\end{eqnarray}
where $\rho=mk_\mu/2\pi^2\hbar^2$ is the density of states at energy $\mu$ and $J=\int d^3\textbf{k}/(2\pi)^3\overline{G}(\textbf{k})\overline{G}^*(\textbf{k})=-i(\ell/v_\mu)(4\pi^3\rho^2\ell/\hbar k_\mu^2)$ is the diagram depicted in Fig. \ref{diagram}b. Now we introduce new variables $\Omega=\epsilon_1-\epsilon_2$, $\Omega^\prime=\epsilon_3-\epsilon_4$, $\epsilon=(\epsilon_1+\epsilon_2)/2$, $\epsilon^\prime=(\epsilon_3+\epsilon_4)/2$, and make use of results of Refs. \cite{Sergey,Henseler}:
\begin{equation}
\int d^3\textbf{r}_3d^3\textbf{r}_4\overline{G^*_{\epsilon_3}(\textbf{r}_3,\textbf{r}^\prime)G^*_{\epsilon_4}(\textbf{r}_4,\textbf{r}^\prime)}\phi(\textbf{r}_3)\phi^*(\textbf{r}_4)=
-\dfrac{2}{\hbar}\int\dfrac{d^3\textbf{k}^\prime}{(2\pi)^3}P_{\epsilon^\prime}(\textbf{r}^\prime,\Omega^\prime)\text{Im}\overline{G}_{\epsilon^\prime}(\textbf{k}^\prime)|\phi(\textbf{k}^\prime)|^2,
\end{equation}
where $P_\epsilon(\textbf{r},\Omega)=\exp(-|\textbf{r}|\sqrt{-i\Omega/\hbar D_\epsilon})/(4\pi D_\epsilon |\textbf{r}|)$ is the Fourier transform of the diffusion propagator at energy $\epsilon$. A similar relation holds for the integrals over $\textbf{r}_1$ and $\textbf{r}_2$.  In the weak disorder limit $k_\mu\ell\gg 1$, $\text{Im}\overline{G}_{\epsilon^\prime}(\textbf{k}^\prime)$ can be approximated by $-\pi\delta(\epsilon^\prime-\epsilon_{k^\prime})$ \cite{Shapiro, Sergey,Henseler}. This finally leads to
\begin{eqnarray}\label{deltapsi_implicit3}
\overline{\Delta n}(\textbf{r},t)&=&
{\rm 2Re}\bigg[\dfrac{gJ}{(2\pi)^3}\dfrac{v_\mu^2}{\rho\hbar\ell^2}\int d^3\textbf{r}^\prime\int\dfrac{d^3\textbf{k}}{(2\pi)^3} \int\dfrac{d^3\textbf{k}^\prime}{(2\pi)^3}\int d\Omega\int d\Omega^\prime
e^{-i\Omega t/\hbar}\times \nonumber\\
&&|\phi(\textbf{k})|^2|\phi(\textbf{k}^\prime)|^2P_{\epsilon_k}(\textbf{r}^\prime,\Omega-\Omega^\prime)P_{\epsilon_{k^\prime}}(\textbf{r}^\prime,\Omega^\prime)P_{\epsilon_k+\Omega^\prime/2}(\textbf{r}-\textbf{r}^\prime,\Omega)
\bigg].
\end{eqnarray}

The main difficulty of Eq. (\ref{deltapsi_implicit3}) lies in the four coupled integrals over energies and momenta, which make the exact calculation of $\overline{\Delta n}$ complicated. However, after some algebra and in the limit of long times, all integrals can be performed. The main lines of this calculation are reported in Appendix \ref{appendix}. We find that for $t\gg t_2=(\ell/v_\mu)(r/\ell)^4(k_\mu\ell)$,  $\overline{\Delta n}$ becomes position independent as $\bar{n}^{(0)}$, and
\begin{equation}\label{result}
\overline{\Delta n}(\textbf{r},t)\simeq -p\dfrac{N}{(D_\mu t)^{3/2}}.
\end{equation}
It is quite remarkable that the first-order correction $\overline{\Delta n}$ has exactly the same time dependence as $\bar{n}^{(0)}$ at long times. Comparison of Eq. (\ref{result}) to $\bar{n}_{\text{dif}}^{(0)}\simeq N/(D_\mu t)^{3/2}$ provides the criterion $p\ll 1$ as a condition of validity of our perturbative approach. As discussed in section \ref{mfp}, this criterion is usually satisfied in typical experiments, which validates the assumption of negligible interactions during expansion made in previous works \cite{Shapiro, Henseler}. 

Few comments are in order. First, the negative sign in Eq. (\ref{result}) indicates that repulsive interactions ($p>0$) tend to reinforce the diffusion process and the contrary for attractive interactions, which might have been expected. This appears more clearly if one writes the total density profile as
\begin{equation}\label{total_n}
\bar{n}(\textbf{r},t)\simeq \dfrac{N}{(D_\text{eff}t)^{3/2}},\ t_2\ll t\ll t_\text{loc},
\end{equation}
where we have defined an effective diffusion coefficient $D_\text{eff}=D_\mu/(1-p)^{2/3}$ taking interactions into account. For repulsive interactions for example, $p>0$ and therefore $D_\text{eff}>D_\mu$. Thus repulsive interactions seem to compete with localization, even though calculations of higher-order diagrams or a complete study of interactions within the localized regime would be necessary to confirm this effect.
For extremely weak disorder ($k_\mu\ell\rightarrow\infty$) $p$ vanishes and $D_\text{eff}\simeq D_\mu$ whatever the interaction strength. 

Another important remark is that the magnitude of the correction term (\ref{result}) can be modified by increasing the strength of disorder independently of the strength of interactions, or vice versa. In particular, if we extrapolate our calculation to $k_\mu\ell\sim 1$, we find that even weak interactions may significantly affect the expansion of the condensate. In this case, the scattering length $a$ must be much less than $\ell/N$ to justify the neglect of interactions. This refines the condition of validity for the results of Ref. \cite{Sergey}, where interactions were assumed weak, but no precise condition of weakness was given.

\begin{figure}[]
\includegraphics[width=12cm]{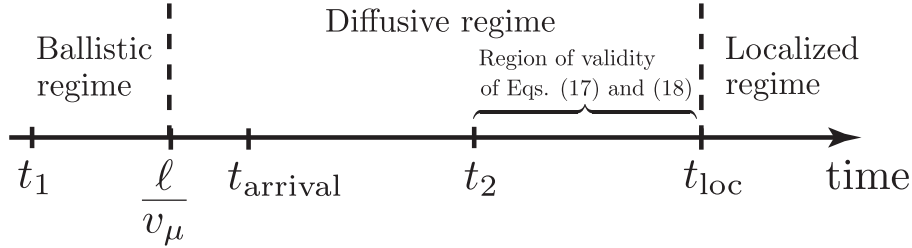}
\caption{\label{Time_scales} 
Ordering of the different time scales that we use throughout the paper. $\ell/v_\mu$ is the mean free time, $t_1$ is the time after which the mean free path is not affected by interactions, $t_{\text{arrival}}$ is the time at which the atomic density free of interactions is maximum, $t_2$ is the limit of long times for nonlinear corrections to the atomic density, and $t_{\text{loc}}$ is the time after which Anderson localization cannot be neglected anymore. In this scheme, the limits of weak disorder $k_\mu\ell\gg 1$ and large distances $r\gg\ell$ are assumed.
}  
\end{figure}

Our result (\ref{result}) is only valid in the time interval $[t_2,t_{\text{loc}}]$. For weak disorder, $t_2\ll t_{\text{loc}}$, which gives a broad range of validity to Eq. (\ref{result}). For the sake of clarity we summarize all the time scales that we introduced above in Fig. \ref{Time_scales}. 

\section{Conclusion}

In conclusion, we developed a perturbative technique to study interaction corrections to the density of a Bose-Einstein condensate expanding in a three-dimensional random potential. Our results apply to the case of weak disorder. We have shown that at the first-order of perturbation theory and at long times, interactions simply renormalize the effective diffusion coefficient of the condensate. Repulsive interactions between atoms accelerate the diffusion process whereas attractive interactions slow it down. In addition, we have found that the effect of interactions on the average density is controlled by a single parameter $p$, depending on both disorder and interaction strengths.  Remarkably, the corrections to the density profile tend to increase when disorder becomes stronger, at a fixed strength of interactions.
Our main result (\ref{result}) has been derived under the assumption of a white-noise, uncorrelated random potential. The generalization to correlated potentials used in experiments is straightforward. For example for a speckle potential of amplitude $V_R$ and standard deviation $\sigma_R$, one simply has to replace the mean free path by $k_\mu\hbar^4/\pi m^2V_R^2\sigma_R^2$ \cite{kuhn07}.
The calculation of higher-order interaction corrections would be of considerable interest in order to validate our main conclusions, even if this rises important yet purely technical difficulties.

\section{Acknowledgements}

S.E.S acknowledges financial support of the French ANR (project No. 06-BLAN-0096 CAROL) and the French Ministry of Education and Research. N. Cherroret thanks the French-German International Research Training Group, and G. Maret for his hospitality in the Fachbereich Physik of the University of Konstanz, where most of this work was carried out. Useful discussions with W. B\"{u}hrer at the early stage of this work are gratefully acknowledged. 

\appendix
\section{}
\label{appendix}

In this appendix we give the main lines of the derivation of Eq. (\ref{result}) from Eq. (\ref{deltapsi_implicit3}). The first step consists of transforming both integrals over $\textbf{k}$ and $\textbf{k}^\prime$ into integrals over energies $\epsilon$ and $\epsilon^\prime$ by making use of $\epsilon=\hbar^2k^2/2m$ and $\epsilon^\prime=\hbar^2k^{\prime2}/2m$. Expressing explicitly the diffusion propagators, Eq. (\ref{deltapsi_implicit3}) can be written as
\begin{eqnarray}\label{appendix1}
\overline{\Delta n}(\textbf{r},t)&=&
{\rm 2Re}\bigg[\dfrac{gJ}{(2\pi)^9}\dfrac{v_\mu^2}{\rho\hbar\ell^2}\left(\dfrac{m}{\hbar^2}\right)^3\int_0^\infty r^{\prime 2}dr^\prime \int_0^\pi\sin(\theta)d\theta \int_0^\infty d\epsilon \int_0^\infty d\epsilon^\prime \int_{-2\epsilon}^{2\epsilon} d\Omega\int_{-2\epsilon^\prime}^{2\epsilon^\prime} d\Omega^\prime\nonumber\\
&&\times\sqrt{\epsilon\epsilon^\prime}
|\phi(\epsilon)|^2|\phi(\epsilon^\prime)|^2\exp\left(-|\textbf{r}-\textbf{r}^\prime|\sqrt{\dfrac{-i\Omega}{\hbar D_{\epsilon+\Omega^\prime/2}}}-r^\prime\left(\sqrt{\dfrac{-i\Omega^\prime}{\hbar D_{\epsilon^\prime}}}+\sqrt{\dfrac{-i(\Omega-\Omega^\prime)}{\hbar D_{\epsilon}}}\right)\right)\nonumber\\
&&\times\dfrac{\exp(-i\Omega t/\hbar)}{r^{\prime^2}|\textbf{r}-\textbf{r}^\prime|D_{\epsilon+\Omega^\prime/2}D_{\epsilon^\prime}D_{\epsilon}}\bigg].
\end{eqnarray}
Here the bounds of integration over $\Omega$ and $\Omega^\prime$ are obtained by noting that $\epsilon\pm\Omega/2$ and $\epsilon^\prime\pm\Omega^\prime/2$ are originally energies (see the main text) and therefore positive quantities. As a consequence, $|\Omega|<2\epsilon$ and $|\Omega^\prime|<2\epsilon^\prime$. The energy distributions $|\phi(\epsilon)|^2$ and $|\phi(\epsilon^\prime)|^2$ are deduced from momenta distributions $|\phi(\textbf{k})|^2$ and $|\phi(\textbf{k}^\prime)|^2$ (we keep the same notation $\phi$ for clarity). We have for example:
\begin{equation}\label{appendix2}
|\phi(\epsilon)|^2=\dfrac{15\pi^2N}{(\sqrt{2}k_\mu)^3}(1-\dfrac{\epsilon}{2\mu})H(1-\dfrac{\epsilon}{2\mu}),
\end{equation}
with $H$ the Heaviside step function. In Eq. (\ref{appendix2}) the prefactor has been obtained by requiring the conservation of the total number of atoms $N=\int d^3\textbf{k}|\phi(\textbf{k})|^2/(2\pi)^3$. The integral over $\theta$ is readily performed:
\begin{equation}
\int_0^\pi\sin(\theta)d\theta\dfrac{e^{-|\textbf{r}-\textbf{r}^\prime|x}}{|\textbf{r}-\textbf{r}^\prime|}=\dfrac{1}{xrr^\prime}\left(e^{-x|r-r^\prime|}-e^{-x(r+r^\prime)}\right),
\end{equation}
where $x=\sqrt{\dfrac{-i\Omega}{\hbar D_{\epsilon+\Omega^\prime/2}}}$. Eq. (\ref{appendix1}) then reduces to:
\begin{eqnarray}\label{appendix3}
\overline{\Delta n}(\textbf{r},t)&=&
{\rm 2Re}\bigg[\dfrac{gJ}{(2\pi)^9}\dfrac{v_\mu^2}{\rho\hbar\ell^2}\left(\dfrac{m}{\hbar^2}\right)^3\int_\Lambda^\infty \dfrac{dr^\prime}{r^\prime} \int_0^\infty d\epsilon \int_0^\infty d\epsilon^\prime \int_{-2\epsilon}^{2\epsilon} d\Omega\int_{-2\epsilon^\prime}^{2\epsilon^\prime} d\Omega^\prime\nonumber\\
&&\times\sqrt{\dfrac{\hbar D_{\epsilon+\Omega^\prime/2}}{-i\Omega}}\sqrt{\epsilon\epsilon^\prime}
|\phi(\epsilon)|^2|\phi(\epsilon^\prime)|^2\exp\left(-r^\prime\left(\sqrt{\dfrac{-i\Omega^\prime}{\hbar D_{\epsilon^\prime}}}+\sqrt{\dfrac{-i(\Omega-\Omega^\prime)}{\hbar D_{\epsilon}}}\right)\right)\nonumber\\
&&\times \left(\exp\left(-|r-r^\prime|\sqrt{\dfrac{-i\Omega}{\hbar D_{\epsilon+\Omega^\prime/2}}}\right)-
\exp\left(-(r+r^\prime)\sqrt{\dfrac{-i\Omega}{\hbar D_{\epsilon+\Omega^\prime/2}}}\right)\right)\nonumber\\
&&\times\dfrac{\exp(-i\Omega t/\hbar)}{rD_{\epsilon+\Omega^\prime/2}D_{\epsilon^\prime}D_{\epsilon}}\bigg],
\end{eqnarray}
where we introduced a lower cutoff $\Lambda\sim\ell$ in the integral over $r^\prime$. This cutoff is needed because of the breakdown of diffusion theory at small length scales. The next step consists of calculating the integral over $r^\prime$. Before doing so it is convenient to introduce the  new dimensionless variables $u=\Omega t/2\hbar$, $v=\Omega^\prime t/2\hbar$, $p=\epsilon t/\hbar$, $q=\epsilon^\prime t/\hbar$, $\tau=k_\mu\ell t/(\ell/v_\mu)$, $\rho=(r/\ell)\sqrt{3k_\mu\ell/\sqrt{2}}$, $\rho^\prime=(r/\ell)\sqrt{3k_\mu\ell/\sqrt{2}}/\tau^{1/4}$ and $\Lambda^\prime=(\Lambda/\ell)\sqrt{3k_\mu\ell/\sqrt{2}}/\tau^{1/4}$. Using $D_\epsilon=(\ell/3)\sqrt{2\epsilon/m}$ we obtain
\begin{eqnarray}\label{appendix4}
\overline{\Delta n}(\textbf{r},t)&=&
{\rm Re}\bigg[-iC\int_{\Lambda^\prime}^\infty \dfrac{d\rho^\prime}{\rho^\prime} \int_0^\tau dp \int_0^\tau dq \int_{-p}^{p} du\int_{-q}^{q} dv \dfrac{1}{\tau^{7/2}}\dfrac{\tau^{1/4}}{\rho}\nonumber\\
&&\times\sqrt{\dfrac{\sqrt{p+v}}{-2iu}}
\left(1-\dfrac{p}{\tau}\right)\left(1-\dfrac{q}{\tau}\right)\exp\left(-\rho^\prime\left(\sqrt{\dfrac{-2iv}{\sqrt{q}}}+\sqrt{\dfrac{-2i(u-v)}{\sqrt{p}}}\right)\right)\nonumber\\
&&\times \left(\exp\left(-\left|\dfrac{\rho}{\tau^{1/4}}-\rho^\prime\right|\sqrt{\dfrac{-2iu}{\sqrt{p+v}}}\right)-
\exp\left(-\left(\dfrac{\rho}{\tau^{1/4}}+\rho^\prime\right)\sqrt{\dfrac{-2iu}{\sqrt{p+v}}}\right)\right)\nonumber\\
&&\times\dfrac{\exp(-2iu)}{\sqrt{p+v}}\bigg],
\end{eqnarray}
where the prefactor $C\simeq(gmN^2k_\mu\ell)/(\hbar^2\ell^4)\simeq (aN^2k_\mu\ell)/\ell^4$. We can now perform the integral over $\rho^\prime$. In the limit $\rho\ll\tau^{1/4}$ which corresponds to long times $t\gg t_2$ ($t_2$ is defined in the main text), this integral is
\begin{eqnarray}\label{appendix5}
&&\int_{\Lambda^\prime}^\infty\dfrac{d\rho^\prime}{\rho^\prime}
\left(\exp\left(-\left|\dfrac{\rho}{\tau^{1/4}}-\rho^\prime\right|\sqrt{\dfrac{-2iu}{\sqrt{p+v}}}\right)-
\exp\left(-\left(\dfrac{\rho}{\tau^{1/4}}+\rho^\prime\right)\sqrt{\dfrac{-2iu}{\sqrt{p+v}}}\right)\right)\nonumber\\
&&\times\exp\left(-\rho^\prime\left(\sqrt{\dfrac{-2iv}{\sqrt{q}}}+\sqrt{\dfrac{-2i(u-v)}{\sqrt{p}}}\right)\right)\nonumber\\
&&\simeq2\left(\dfrac{\rho}{\tau^{1/4}}-\Lambda^\prime\right)\sqrt{\dfrac{-2iu}{\sqrt{p+v}}}\simeq2\dfrac{\rho}{\tau^{1/4}}\sqrt{\dfrac{-2iu}{\sqrt{p+v}}},
\end{eqnarray}
where the last equality results from $\rho/\tau^{1/4}\gg\Lambda^\prime$. Inserting Eq. (\ref{appendix5}) into Eq. (\ref{appendix4}) we obtain
\begin{equation}\label{appendix6}
\overline{\Delta n}(\textbf{r},t)=
{\rm 2Re}\left[-i\dfrac{C}{\tau^{7/2}} \int_0^\tau dp \int_0^\tau dq \int_{-p}^{p} du\int_{-q}^{q} dv \dfrac{\exp(-2iu)}{\sqrt{p+v}}\right].
\end{equation}
The four remaining integrals can be readily performed. We finally obtain:
\begin{equation}\label{appendix7}
\overline{\Delta n}(\textbf{r},t)\simeq
\dfrac{-C}{\tau^{3/2}}=-\dfrac{aN^2}{\ell\sqrt{k_\mu\ell}}\dfrac{1}{(D_\mu t)^{3/2}},
\end{equation}
which is Eq. (\ref{result}) of the main text.


\begin{thebibliography}{99}

\bibitem{reviewBEC} For a review see L. Fallani, C. Fort, M. Inguscio. Adv. At. Mol. Opt. Phys. 
\textbf{56}, 119 (2008).

\bibitem{Fallani07} L. Fallani, J. E. Lye, V. Guarrera, C. Fort and M. Inguscio, Phys. Rev. Lett.
\textbf{98}, 130404 (2007).

\bibitem{experimental_papers} J.E. Lye  \emph{et al.}, Phys. Rev. Lett.
\textbf{95}, 070401 (2005); D. Clement \emph{et al.}, \emph{ibid}
\textbf{95}, 170409 (2005); C. Fort \emph{et al.}, \emph{ibid}
\textbf{95}, 170410 (2005). 

\bibitem{theoretical_papers} M. Modugno, Phys. Rev. A \textbf{73}, 013606 (2006); E. Akkermans, S. Ghosh and Z. H. Musslimani, J. Phys. B: At. Mol. Opt. Phys. \textbf{41}, 045302 (2008).

\bibitem{Sanchez-Palencia} L. Sanchez-Palencia \emph{et al.}, Phys. Rev. Lett.
\textbf{98}, 210401 (2007).

\bibitem{Billy} J. Billy \emph{et al.}, Nature
\textbf{453}, 891 (2008).

\bibitem{Roati} G. Roati \emph{et al.}, Nature
\textbf{453}, 895 (2008).

\bibitem{Anderson} P. W. Anderson, Phys. Rev.
\textbf{109}, 1492 (1958).

\bibitem{Sergey} S.E. Skipetrov, A. Minguzzi, B.A. van Tiggelen, and B. Shapiro, Phys. Rev. Lett.
\textbf{100}, 165301 (2008).

\bibitem{Shapiro} B. Shapiro, Phys. Rev. Lett.
\textbf{99}, 060602 (2007).

\bibitem{Pitaevskii} L. Pitaevskii and S. Stringari, \emph{Bose-Einstein Condensation} (Clarendon, Oxford, 2003).

\bibitem{Pikivsky08} A. S. Pikovsky and D. L. Shepelyansky, Phys. Rev. Lett.
\textbf{100}, 094101 (2008). S. Flach, D. O. Krimer, and Ch. Skokos, Phys. Rev. Lett.
\textbf{102}, 024101 (2009).

\bibitem{Akkermans} E. Akkermans and G. Montambaux, \emph{Mesoscopic Physics of Electrons and Photons} (Cambridge University Press, Cambridge, 2007).

\bibitem{Kagan} Yu. Kagan, E.L. Surkov, and G.V. Shlyapnikov, Phys. Rev. A
\textbf{54}, R1753 (1996);
Y. Castin and R. Dum, Phys. Rev. Lett.
\textbf{77}, 5315 (1996).

\bibitem{Spivak} B. Spivak and A. Zyuzin, Phys. Rev. Lett.
\textbf{84}, 1970 (2000).

\bibitem{Henseler} P. Henseler and B. Shapiro, Phys. Rev. A
\textbf{77}, 033624 (2008).

\bibitem{Nicolas08} N. Cherroret and S. E. Skipetrov, Phys. Rev. Lett.
\textbf{101}, 190406 (2008).

\bibitem{Agranovich91} V. M. Agranovich and  V. E. Kravtsov, Phys. Rev. B
\textbf{43}, 13691 (1991); A. Heiderich, R. Maynard and B. A. van Tiggelen, \emph{Opt. Commun.}
\textbf{115}, 392 (1995); T. Wellens and B. Gr\'{e}maud, J. Phys. B \textbf{39}, 4719 (2006).

\bibitem{kuhn07}
R.C. Kuhn \emph{et al.},
New. J. Phys. \textbf{9}, 161 (2007).

\end{thebibliography}
\end{document}